\documentclass[twocolumn,10pt]{article}
\usepackage[utf8]{inputenc}
\usepackage{mathptmx}
\usepackage[affil-it]{authblk} 
\usepackage[a4paper, margin=1in]{geometry}
\usepackage{sectsty}
\sectionfont{\fontsize{10}{10}\selectfont}
\subsectionfont{\fontsize{10}{10}\selectfont}
\usepackage{titlesec}
\titleformat*{\subsection}{\normalsize\itshape}
\usepackage{indentfirst}
\usepackage{array,multirow}
\usepackage{fancyhdr}
\usepackage{graphicx}
\usepackage{caption}
\usepackage{float}
\usepackage{hyperref}
\newcommand{\ctext}[1]{\raise0.2ex\hbox{\textcircled{\scriptsize{#1}}}}

\title{\LARGE \bf
What Is the Gaze Behavior of Pedestrians\\in Interactions with an Automated Vehicle\\When They Do Not Understand Its Intentions?
}
\author{
        \large \textbf{Hailong Liu~$^{*}$, Takatsugu Hirayama~$^{\#}$, Luis Yoichi Morales~$^{\#}$, Hiroshi Murase~$^{*}$}\\  
        $^{*}$~Graduate School of Informatics, Nagoya University\\
        $^{\#}$~Institutes of Innovation for Future Society, Nagoya University\\
        \normalsize
        Furo-cho, Chikusa-ku, Nagoya, Aichi, 464-8601, JAPAN\\
        E-mail: lhl881210@live.com
        }
\date{}

\begin{document}
\thispagestyle{empty}
\pagestyle{empty}
\maketitle
\begin{abstract}
Interactions between pedestrians and automated vehicles~(AVs) will increase significantly with the popularity of AV. However, pedestrians often have not enough trust on the AVs , particularly when they are confused about an AV's intention in a interaction.
This study seeks to evaluate if pedestrians clearly understand the driving intentions of AVs in interactions and presents experimental research on the relationship between gaze behaviors of pedestrians and their understanding of the intentions of the AV.
The hypothesis investigated in this study was that the less the pedestrian understands the driving intentions of the AV, the longer the duration of their gazing behavior will be.
A pedestrian--vehicle interaction experiment was designed to verify the proposed hypothesis.
A robotic wheelchair was used as the manual driving vehicle~(MV) and AV for interacting with pedestrians while pedestrians' gaze data and their subjective evaluation of the driving intentions were recorded.
The experimental results supported our hypothesis as there was a negative correlation between the pedestrians' gaze duration on the AV and their understanding of the driving intentions of the AV.
Moreover, the gaze duration of most of the pedestrians on the MV was shorter than that on an AV.
Therefore, we conclude with two recommendations to designers of external human-machine interfaces (eHMI):
(1) when a pedestrian is engaged in an interaction with an AV, the driving intentions of the AV should be provided;
(2) if the pedestrian still gazes at the AV after the AV displays its driving intentions, the AV should provide clearer information about its driving intentions.
\end{abstract}

\section{INTRODUCTION}

\begin{figure}[tb]
\centering
\includegraphics[clip=true,trim=0.0in 0.0in 0.0in 0.1in,width=1\linewidth]{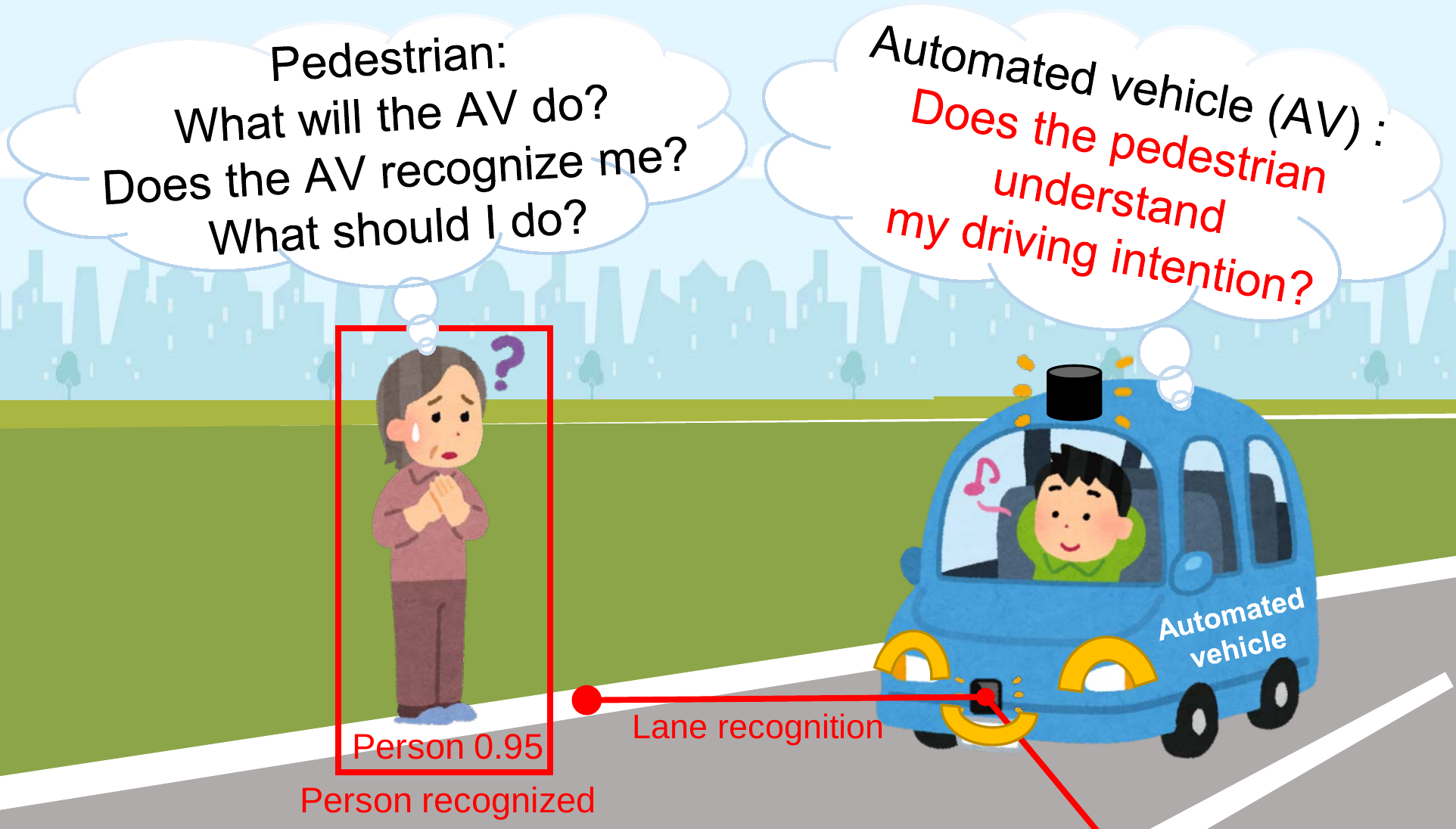}
\caption{Pedestrian does not understand the intention of the AV during the interaction; therefore, she does not trust it. The AV cannot determine if the pedestrian understood its driving intentions.}
 \label{fig:prob}
 \vspace{-5mm}
\end{figure}

The functions and capabilities of automated driving system~(ADS) are continuously being improved with increasing emphasis on safety.
The ADS is not only used in ordinary cars, but it is also widely applied to personal mobility vehicles~(PMVs)~\cite{watanabe2015,andersen2016autonomous} and delivery robots~\cite{hoffmann2018regulatory,jennings2019study}.
These vehicles equipped with ADS are called automated vehicles (AVs).
The number of interactions between AVs and pedestrians will increase significantly with the popularity of AVs, since some PMVs and delivery robots interact with pedestrians in shared spaces and sidewalks~\cite{morales2018personal}.
However, AVs have encountered some resistance during the stage of popularization in society~\cite{epprecht2014anticipating,Hancock2019}.
One aspect of the resistance is that the public do not trust the AVs~\cite{Pettigrew2019} when they do not understand the capabilities and performance of the AVs in an interaction, particularly when the intention of the AV is not clear to them.
A typical example of such interaction is depicted in Fig.~\ref{fig:prob}: a pedestrian may feel worried and overwhelmed when she wish to cross the road in front of an AV because it is difficult to determine whether the AV has detected her, could keep her safe, and will give her the right-of-way.
Furthermore, the same problem is likely to occur when a pedestrian interacts with AVs in shared spaces and sidewalks.
To address this issue, Habibovic et al. and Rasouli et al. concluded that it is important for vehicles to communicate with pedestrians about their driving intentions during the interaction~\cite{Habibovic2018,rasouli2019}.
To realize a communication between pedestrians and AVs, the external human-machine interface~(eHMI) for automated vehicles could be considered a useful method~\cite{keferbock2015strategies,Burns2019}.

Using the eHMI, the AV can clearly and quickly convey the driving intentions of the AV to the pedestrians thus improving the acceptability and the popularization of AVs.
However, as illustrated in the Fig.~\ref{fig:prob},
situational and timing details of providing information on driving intentions of AVs to pedestrians and evaluation of the pedestrians' current understanding of the driving intentions of the AV remain undetermined.
Therefore, to evaluate whether pedestrians clearly understand driving intentions of AVs during interactions is the research objective of this study.

The main purpose of this paper is to find an objective measure to evaluate the uncertainty of pedestrians when they try to understand the driving intentions of the AV.
For this task, we surveyed the literature on cognitive psychology.
Many studies reported that gaze duration is related to the levels of understanding~\cite{LIVERSEDGE199855,Rayner2006,Gomez2014,Sanches2017,Sanches2018}.
Therefore, the gaze duration of pedestrians on the AV is the focus of this study.
As a further consideration in the pedestrians' situation awareness process, we consider that if pedestrians have an unclear understanding of the AV's driving intentions, then they will continue to gaze at the AV.
Thus, the gaze duration of pedestrians on an AV could be used to represent the degree of the pedestrian's understanding of the AV's driving intentions.
The experiment conducted in this study aims to verify the correlation between the pedestrians' gaze duration on an AV and their understanding of the AV's driving intentions during the interaction.

\section{RELATED WORKS}
Many subjective evaluation methods have been used to obtain pedestrians' views on participant-AV interactions~\cite{watanabe2015,Stefanie2020,Clercq2019,Lee2019,Locken2019}.
To establish communication between an AV and a pedestrian, Watanabe et al. mounted a projector on a robotic wheelchair to show the navigational intention of the wheelchair.
Then they investigated pedestrians' understanding of the driving intentions and the feeling of comfort in the interaction with a questionnaire~\cite{watanabe2015}.
Stefanie et al. evaluated the efficacy of eHMIs, such as a steady, flashing, and sweeping light signal to communicate an AV's intention to yield, by using questionnaires and structured interviews with participants~\cite{Stefanie2020}.
Clercq et al. compared pedestrians' understanding of information from different types of eHMI, such as front brake lights, knight-rider animation, smiley face, and text.
Participants continuously evaluated their feeling of safety by pressing a button during the AV-pedestrian interaction~\cite{Clercq2019}.
Similarly, Lee et al. and L\"{o}cken et al. analyzed the difference of pedestrians' subjective evaluations for various types of eHMI in a virtual reality space using a questionnaire~\cite{Lee2019,Locken2019}.

Meanwhile, objective variables have been used to analyze the interaction between pedestrians and vehicles in a small number of studies~\cite{dey2019gaze,Fuest2018,Fuest2020}.
Dey et al. conducted an eye-tracking study with 26 participants in a road-crossing situation.
They found that the gaze of the pedestrians gradually moved to the windshield at the driver's position as the manual driving vehicle (MV)
was approaching~\cite{dey2019gaze}.
Thus, the pedestrians look at the driver and hope to acquire the driver's intention when they cannot recognize whether the vehicle will stop.
However, for level 3-5 AVs~\cite{SAE_j3016_2016}, pedestrians cannot understand the driving intentions from the driver or the passenger because the AV usually does not require the operations of a driver.
Therefore, pedestrians can only recognize the AV's intention by the driving behaviors, such as speed, acceleration, and direction.
To investigate whether pedestrians can recognize driving intentions from AV driving behavior, Fuest et al. focused on the intention recognition time when AVs interacted with pedestrians~\cite{Fuest2018,Fuest2020}.
To observe the intention recognition time, the participants were asked to press a button when they thought they recognized the vehicle's intention.
The study compared the participants' intention recognition time for the AV and the participants' subjective evaluation results of understanding the AV's driving intentions in different right-of-way situations.
However, it did not examine the relationship between the intention recognition time and an understanding of the AV's driving intentions.

\section{HYPOTHESIS}
\begin{figure*}[tb]
\centering
\includegraphics[width=0.75\linewidth]{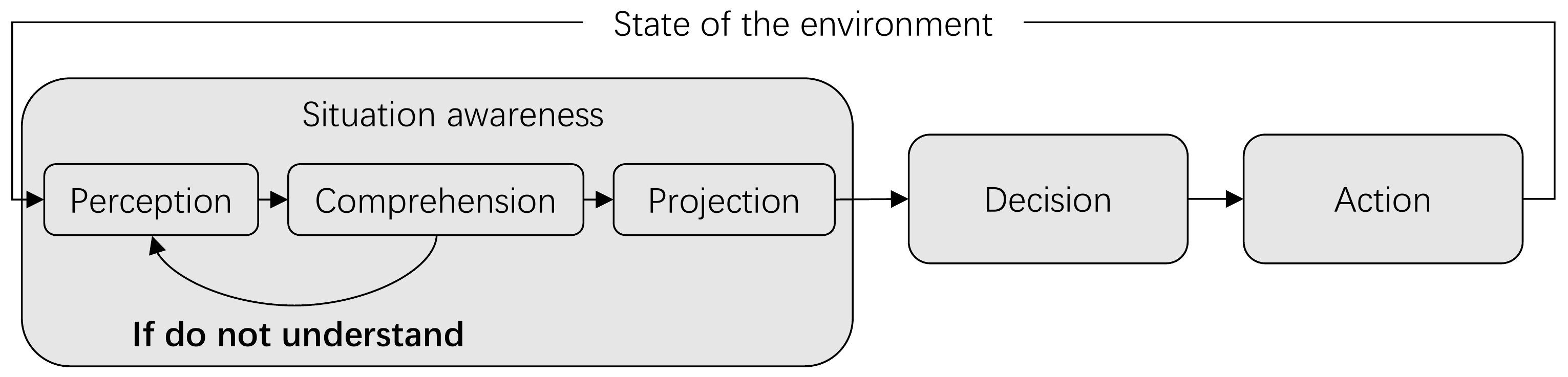}
\caption{The decision-making model of a pedestrian has three steps: situation awareness, decision, and action. The hypothesis of this work is that pedestrians will repeatedly observe the vehicle when they are not certain of the vehicle's driving intention.}
\label{fig:model}
\vspace{-5mm}
\end{figure*}

The main purpose of this paper is to find an objective measure to represent the uncertainty of pedestrians when they try to understand the driving intentions of the AV.

In our model, the pedestrians use their cognitive abilities to assess the situation within their surrounding environment that includes the AV. 
Then they acquire situation awareness, which is used to make a decision.
Finally, the pedestrians act according to the decision.
The above process is represented in Fig.~\ref{fig:model}.
This study focuses on situation awareness, which includes three steps: perception, comprehension, and projection~\cite{endsley2017toward}.

Firstly, situation awareness relies on the perception of things in the surrounding environment, such as the AV's relative position, distance, and speed.
Secondly, comprehension is taken as the understanding of the current state of the AV, such as the driving intention of the AV.
Thirdly, based on the result of the comprehension, the pedestrian will predict the driving behavior and moving trajectory of the AV.
Based on the described process, we suppose that if the pedestrian does not understand the AV's driving intentions clearly in the comprehension step, the pedestrian will return to the perception step to observe the vehicle until they believe they have correctly understood the driving intentions of the AV.
In summary, the gaze duration could express pedestrians' requirements to understand the driving intentions of an AV while also represent a degree of the pedestrian's understanding of the AV's driving intentions.

Therefore, we propose a hypothesis based on the above process:
\textbf{
the less the pedestrian understands the driving intentions of the AV, the longer the duration of their gazing behavior will be.}

\section{PEDESTRIAN--VEHICLE INTERACTION EXPERIMENT}

To verify the proposed hypothesis, we assumed that a low speed AV, a PMV, would interact with pedestrians in a shared space.
We used a robotic wheelchair as the AV that could automatically drive following predesigned routes, but it did not have the functionality to interact with pedestrians. 
Therefore, we conducted a Wizard of Oz (WOZ) experiment: an experiment in which the participants are led to believe they are interacting with an autonomous system but the system is operated or partially operated by the experimenter. 
WOZ experiments are often used in the research of human–machine interaction.
Throughout our experiment, an experimenter secretly controlled the AV and was able to stop it according to the actual situation during the participant--AV interactions.

The goal of this experiment was to record the gaze duration of the participants when they interacted with a manual driving vehicle~(MV) and an AV.
Meanwhile, the participants' subjective evaluation of the interactions was recorded using questionnaires.
In addition, the correlation between the participants' subjective evaluation and gaze duration on the vehicle was analyzed.

This experiment was approved by the ethics review committee of Institute of Innovation for Future Society, Nagoya University.

\subsection{Experimental condition}

\begin{figure}[tb]
\centering
\includegraphics[width=\linewidth]{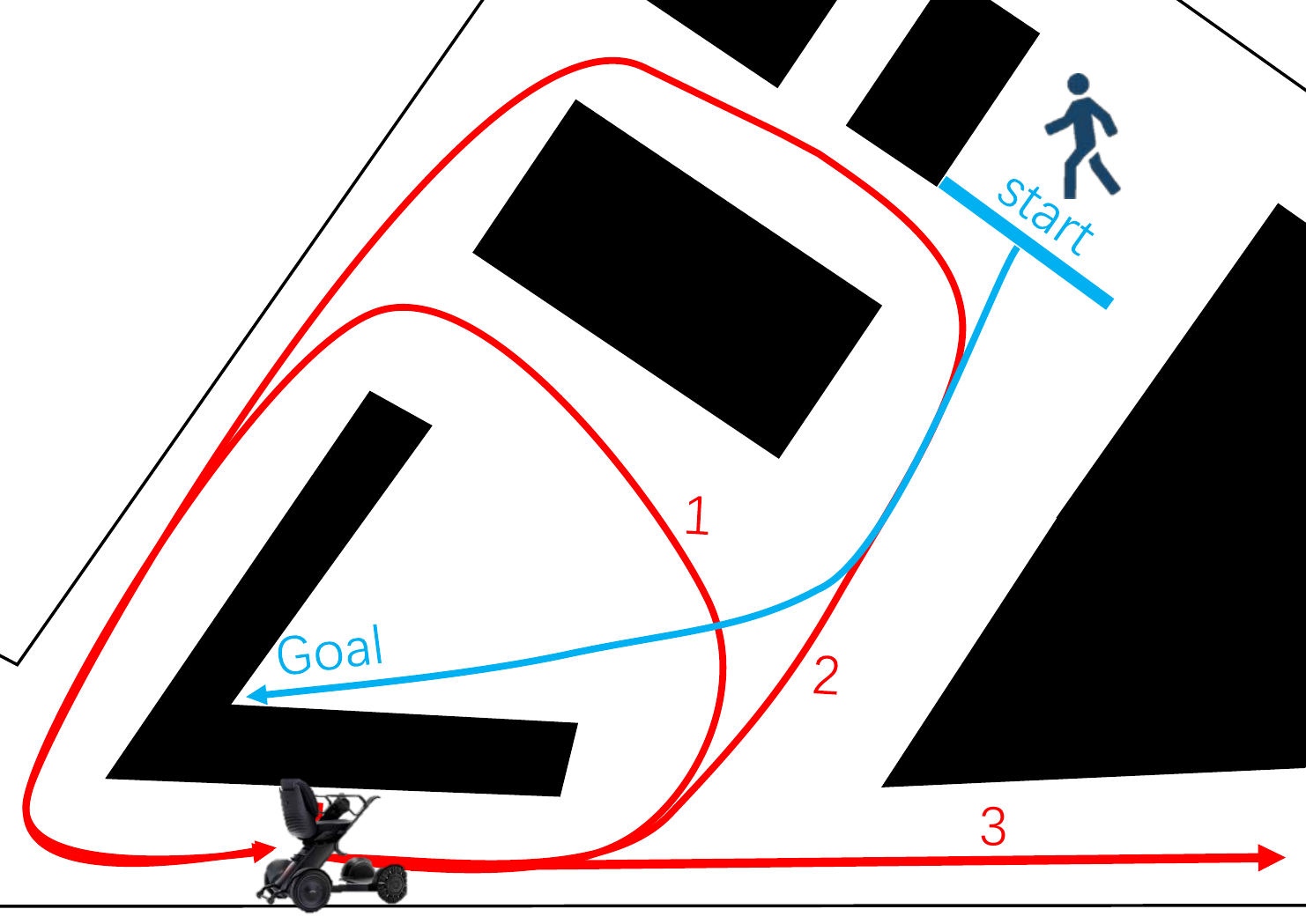}
\caption{Three designed routes followed by the robotic wheelchair in red and one pedestrian route in blue. The black areas are obstacles that cannot be passed.}
\label{fig:map}
\vspace{3mm}
\centering
\includegraphics[width=1\linewidth]{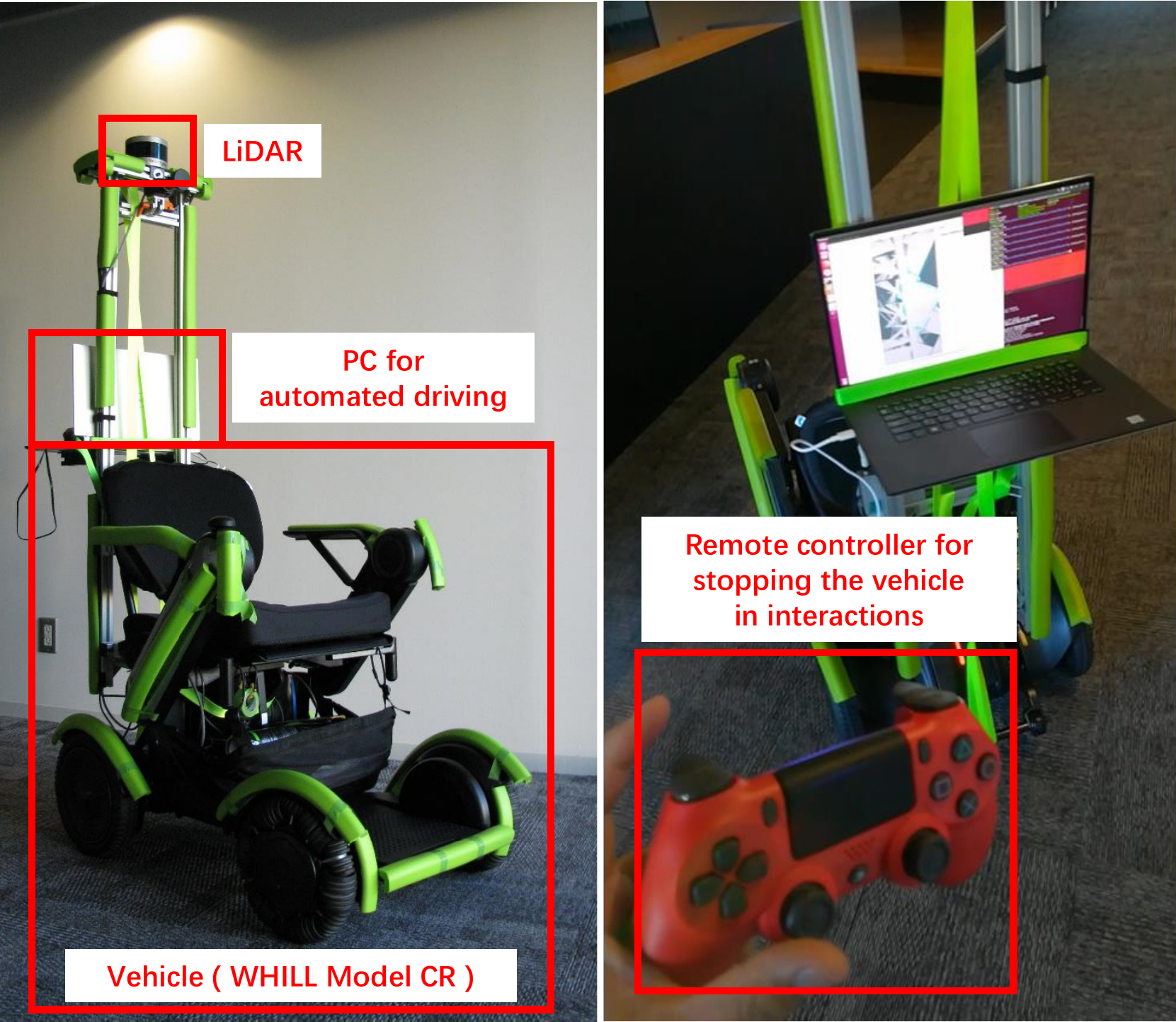}
\caption{The robotic wheelchair equipped with a LiDAR, wheel encoders, and a wireless remote controller for the automated driving.}
\label{fig:whill}
\vspace{-6mm}
\end{figure}

\subsubsection{Movement routes}
The movement routes for pedestrian--vehicle interactions are shown in Fig~\ref{fig:map}.
A robotic wheelchair equipped with an automated driving system acted as the experimental vehicle.
It did not have the function to interact with pedestrians, but it could automatically drive following predesigned routes.
Three driving routes were prepared in advance, and they are shown in Fig.~\ref{fig:map} with red lines.
The participants were asked to walk at their normal pace from the starting point to the end point that is shown in Fig.~\ref{fig:map} with a blue line.
They interacted with the vehicle when the vehicle was on route 1 or 2.
The interaction when the AV takes route 1 simulates a scene of a crossing paths.
The interaction when the AV takes route 2 simulates a scene in which a pedestrian and an AV approach each other on a straight road.
Although the pedestrians did not interact with the vehicle on route 3, they could still observe, evaluate, and predict the behavior of the vehicle.

\subsubsection{Vehicle}

A robotic wheelchair--{\it WHILL Model CR}, shown in the left part of Fig.~\ref{fig:whill}, was used as the experimental vehicle,.
For manual driving, an operator rode on the robotic wheelchair~(i.e. MV) and used the available joystick to manipulate it.
The operator did not actively convey information about driving intentions through actions or language to the participants.
Meanwhile, there were no turn lights or brake lights on the vehicle to communicate the status of the vehicle to the participants.
For automated driving, the robotic wheelchair~(i.e. AV) was equipped with a multilayered lidar (Velodyne VLP--16) and wheel encoders.
The lidar was utilized for self-localization on a previously built environmental map. 
We recorded the time-stamped position of the vehicle and its linear and angular velocities and accelerations.
The AV had an automatic brake function that was applied when there was an obstacle within 0.5 meters directly in front of it.
This function was used to ensure the safety of the participants in the experiment.
The AV could automatically drive following the predesigned driving routes, but it did not have the functions to recognize pedestrians and interact with them, e.g. automatically yielding the right--of--ways.
The experimenter used a wireless remote controller, which is shown in the right part of Fig.~\ref{fig:whill}, to secretly control the AV and to stop it according to the actual situation during participant--AV interactions.
The maximum speed of the vehicle was limited to 1 [m/s] for automated and manual driving.
The driving route was randomly selected before the experiment.
During the experiment, no devices were showing information about the driving status and intentions of the AV to the participants.

\subsubsection{Participants}
Ten participants, three females and seven males,
aged 20-29 (mean: 23.2, std 2.14) were invited to this experiment.
Nine participants had a driver’s license, and one participant did not.
None of them had any prior experience with AVs.

To measure the gaze behavior of the participants during walking, they were asked to use a wearable eye tracker--{\it Tobii Pro Glasses 2} throughout the experiment.

\subsection{Experimental procedure}
First, we explained to the participants following information before the experiment:
\begin{enumerate}
\item Please walk at your normal pace from the starting point to the end point.
\item An experimental vehicle  will interact with you during your walk.
\item The vehicle has three driving routes, and it will randomly select a route before each trial.
\item The vehicle will be used for manual and fully automated driving.
\item An experimenter will ride on the vehicle when driving manually, and no one will ride on the vehicle when it is driving autonomously.
\item The maximum speed of the vehicle is limited to 1 [m/s] for both manual and automated driving.
\item During automated driving, the vehicle's sensors can recognize the surrounding objects and the pedestrian. As a result, the vehicle will automatically determine the driving behavior when interacting with the pedestrian.~(\textbf{False information})
\item There is no perfect system in the world, so this experiment still involves some risks.
\item If you think the behavior of the AV threatens you, please stay away from it.
\item When the interaction is over, the experimenter will use a wireless remote controller to manually control the AV so it returns to its starting position.
\end{enumerate}

\begin{figure}[h!tb]
\vspace{3mm}
\centering
\includegraphics[clip=true,trim=0mm 30mm 0mm 30mm, width=1\linewidth]{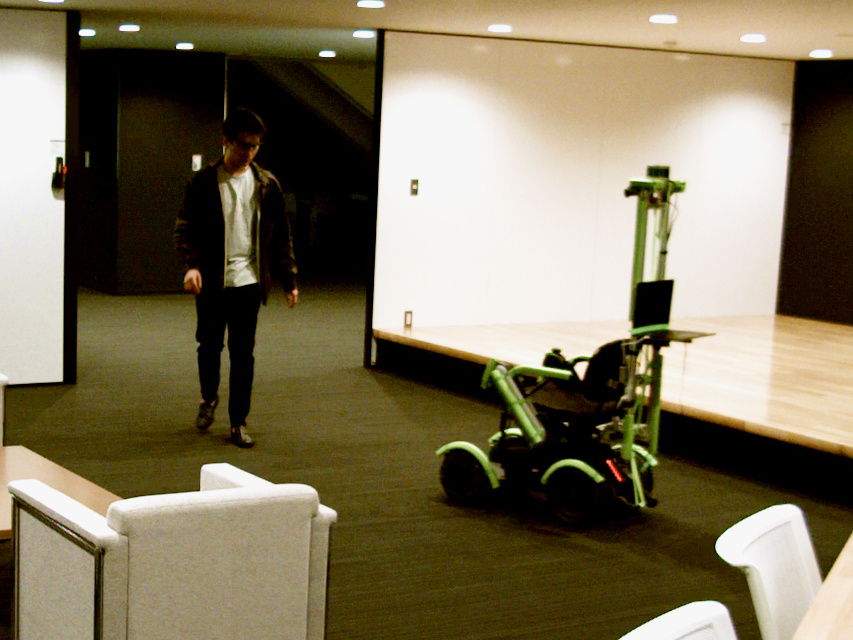}
\vspace{0.5mm}
\caption{Scene of an interaction between a participant and the AV during the experiment.}
\label{fig:img_inter}
\vspace{6mm}
\centering
\includegraphics[width=1\linewidth]{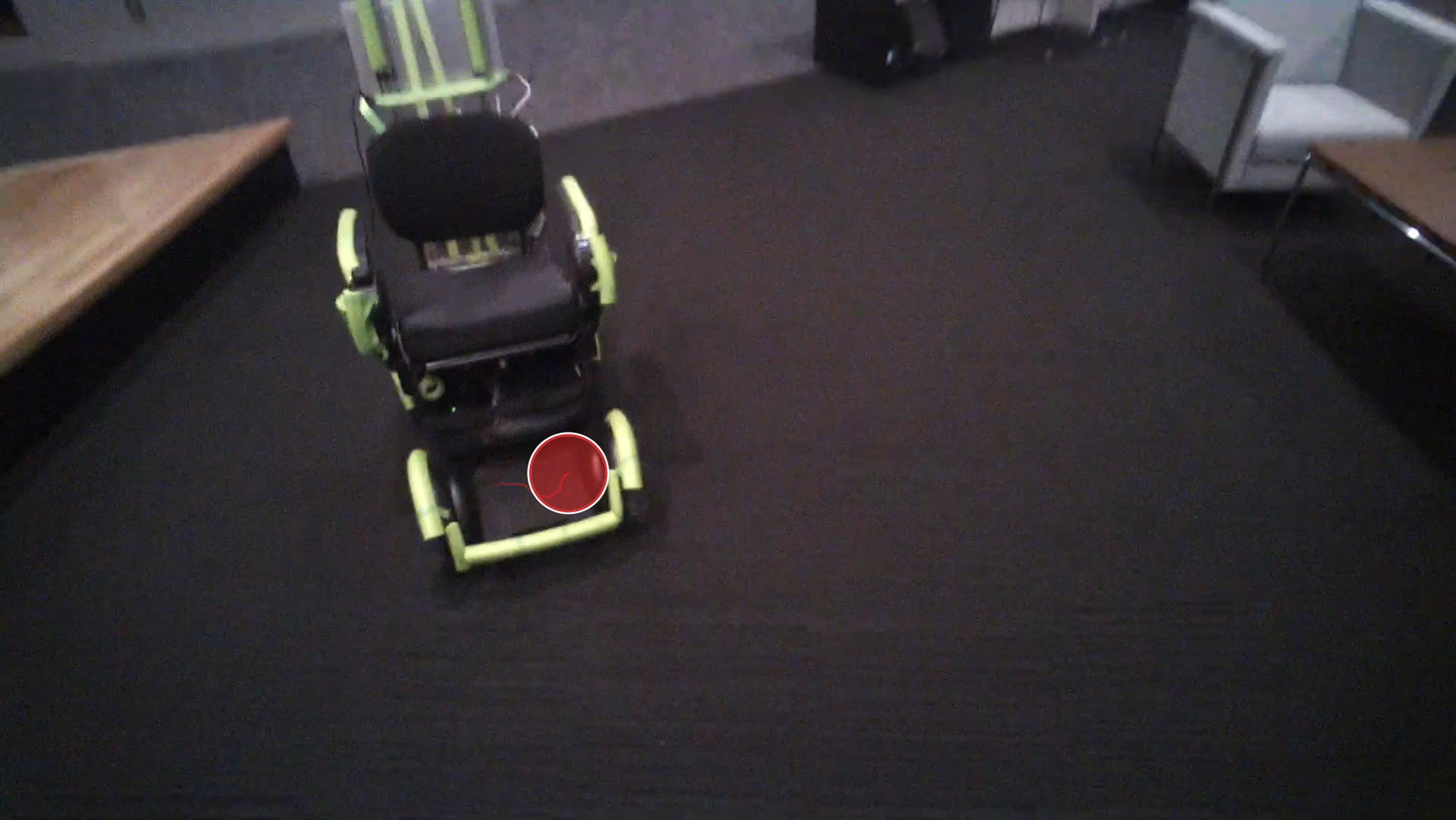}
\vspace{0.5mm}
\caption{Participant gaze view recorded by {\it Tobii Pro Glasses 2} eye-tracking device. The red dot shows the region of interest of the participant.}
\label{fig:img_view}
\vspace{6mm}
\centering
\includegraphics[width=1\linewidth]{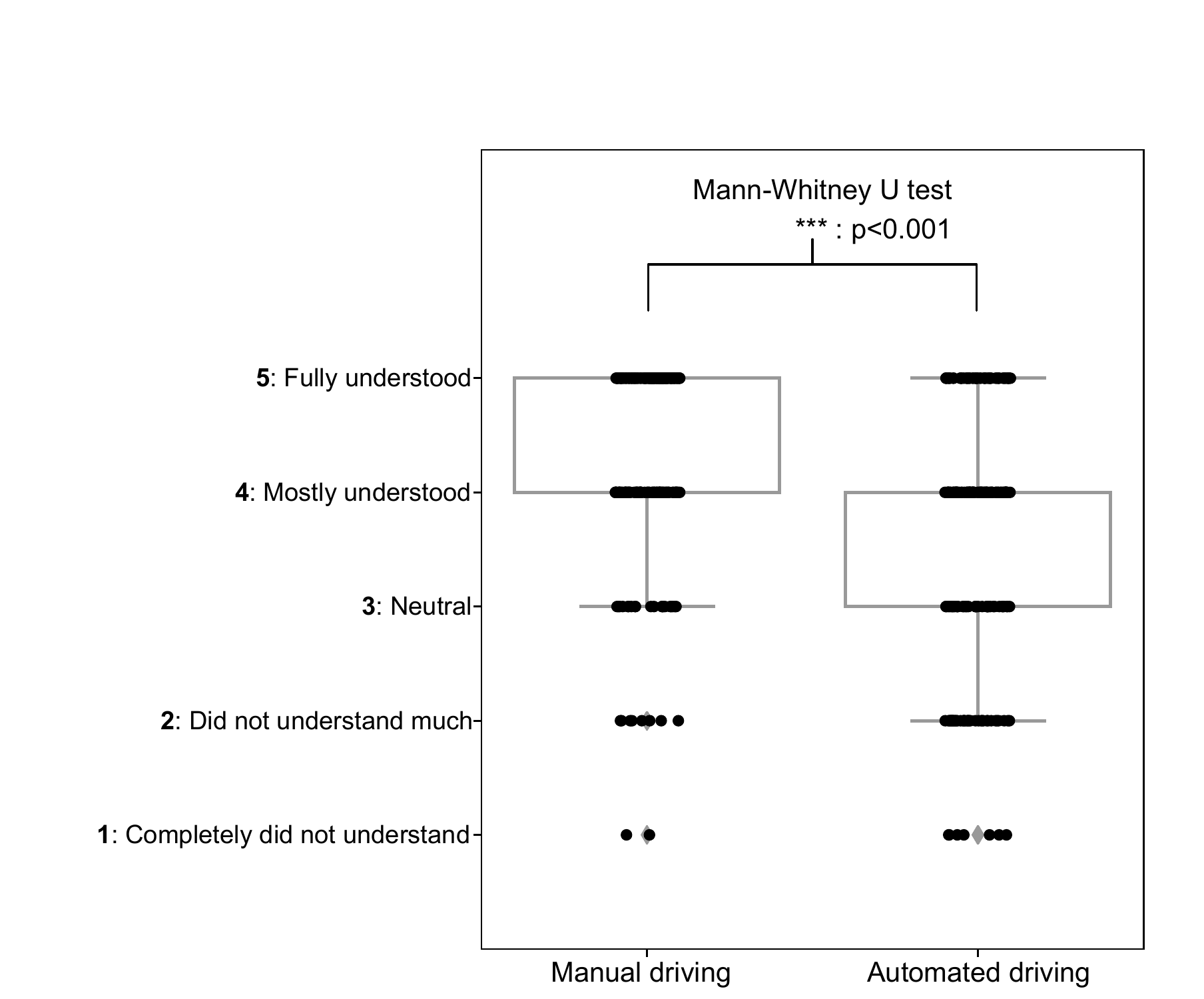}
\vspace{0.5mm}
\caption{ Evaluation of the understanding of driving intentions of the MV and AV by the ten participants.
}
\label{fig:evaluation}
\end{figure}

Next, the participants were asked to interact with the MV for 20 trials.
Then the participants were asked to interact with the AV for 20 trials.
This experimental sequence was designed to take into account the order--dependence of MV and AV
because AVs have not been popular, but MVs have been regularly used for many years.
In other words, the participants were used to interacting with MVs; however, interacting with an AV was a new concept.
An example interaction between the participant and the AV in the experiment is shown in Fig.~\ref{fig:img_inter}.
This figure illustrates that the participant was observing the vehicle in an interaction.
A field of view from the participant is presented in Fig.~\ref{fig:img_view}.
The red circle in Fig.~\ref{fig:img_view} represents the participant's gaze area and the red line represents the movement of the gaze area.

After each trial of interactions, the participants were asked to subjectively evaluate their understanding of the driving intentions on a five-level scale in Japanese, with the levels as follows:
\begin{enumerate}
\item[\ctext{1}] Completely did not understand; 
\item[\ctext{2}] Did not understand much; \item[\ctext{3}] Neutral;
\item[\ctext{4}] Mostly understood;
\item[\ctext{5}] Fully understood.
\end{enumerate}

Finally, an unstructured interview was conducted with each participant in 30$\sim$40 minutes after the interactions. 
The interview mainly listened to the participants' during the experiment and impressions of interacting with the AV.

\section{EXPERIMENTAL RESULTS AND DISCUSSION}
In this section, the results of the subjective evaluation, the gaze duration, and the relation between them are explained.

In the experiment, we found that some participants' long eyelashes affected the collected gaze data and created noise.
To reduce the influence of the noise, some participants were asked to participate in one more interaction with the vehicles: participant $\#9$ with the MV and participants $\#5$,$\#7$,$\#9$,$\#10$ with the AV.
The experimental conditions of the added trials were the same as the previous experimental conditions.
Besides, the gaze data of participant $\#3$ from the first three interactions with the MV were removed due to equipment problems.
In total, gaze durations and subjective evaluations of ten participants were measured in \textbf{198 interactions with the MV} and \textbf{204 interactions with the AV}.

\subsection{Subjective evaluation of the understanding of driving intentions}

Results of the evaluation of the understanding of driving intentions by the participants are shown in Fig.~\ref{fig:evaluation}.
Participants mostly evaluated the MV's driving intentions with {\it \ctext{4} Mostly understood} and {\it \ctext{5} Fully understood} as shown in separate box plots of Fig.~\ref{fig:evaluation}.
Participants mostly evaluated the AV's driving intentions as {\it \ctext{3} Neutral} and {\it \ctext{4} Mostly understood}.
Additionally, participants rated AV's intentions as {\it \ctext{1} Completely did not understand} and {\it \ctext{2} Did not understand much}  more times than the MV’s intentions.

Furthermore, a statistical hypothesis test was performed for the subjective evaluation results.
The null hypothesis of this statistical hypothesis states that ``there was no significant difference in the subjective evaluation results of the understanding of the driving intentions of the MV and AV.''
The Mann--Whitney U test~\cite{McKnight2010}, which is a non-parametric test method, was used.
There were three reasons to use Mann--Whitney U test.
The first reason was that the trials of pedestrian interactions with the MV and AV occurred independently.
The second reason was that the results of the five levels of evaluation were considered to have a multinomial distribution instead of a Gaussian distribution.
The third reason was that the numbers of trials of MV and AV interactions were not the same because of noise and equipment problems.
The p-value of the statistical hypothesis testing result was smaller than $0.001$.
Therefore, the null hypothesis was rejected, and the results show that the ten participants perceived the driving intentions of the AV were more difficult to understand during the interaction than those of the MV.

\subsection{Gaze durations on the MV and AV}
\begin{figure*}[tb]
\centering
\includegraphics[width=1\linewidth]{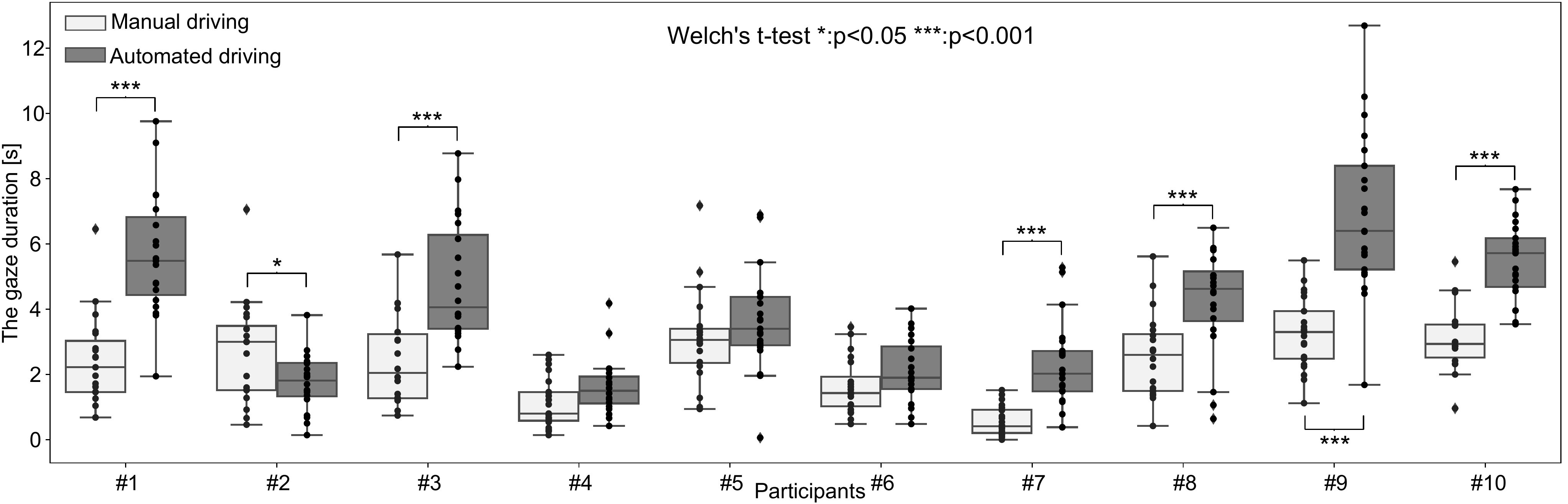}
\vspace{-6mm}
\caption{Comparison of gaze durations of the ten participants interacting with the MV and AV. Light gray boxes and dark gray boxes show the gaze durations on the MV and AV, respectively.}
\label{fig:time}
\vspace{-1mm}
\end{figure*}

Gaze durations of each participant were analyzed from the measured gaze data. 
A gaze duration is defined as the cumulative time during which a participant gazed at the vehicle in a trial. 
The gaze durations of the ten participants are presented in Fig.~\ref{fig:time}.
From the above data, two results were determined. 
The first result is that the length of the gaze durations of each participant was different.
This indicates that everyone has their own tendencies of observing things.
The second result is that most of the participants' gaze durations on the AV were longer than on the MV, except for participant $\#2$.

The significant difference between the gaze durations on the MV and AV was tested independently for each participant.
Welch's t--test~\cite{delacre2017psychologists} was used for this statistical hypothesis test because the distribution of the gaze durations was assumed to have a Gaussian distribution. 
The variance of the gaze durations was different for the AV and MV.
The null hypothesis states that ``there was no significant difference in the participant's gaze duration on the MV and AV.''
The results of this statistical hypothesis test are shown in Fig.~\ref{fig:time}.
For participants $\#1$, $\#3$, $\#7$, $\#8$, $\#9$, and $\#10$, the p--values were lower than $0.001$.
Meanwhile, for participant $\#2$, the p--value was lower than $0.05$.
The null hypotheses for those participants were rejected.
Thus, their gaze durations on the MV and AV were significantly different.
However, the null hypotheses for participants $\#4$, $\#5$, and $\#6$ were accepted.

\subsection{Relationship between the understanding of driving intentions and gaze durations}
\begin{figure*}[tb]
\begin{minipage}{0.48\linewidth}
\centering
\includegraphics[width=1\linewidth]{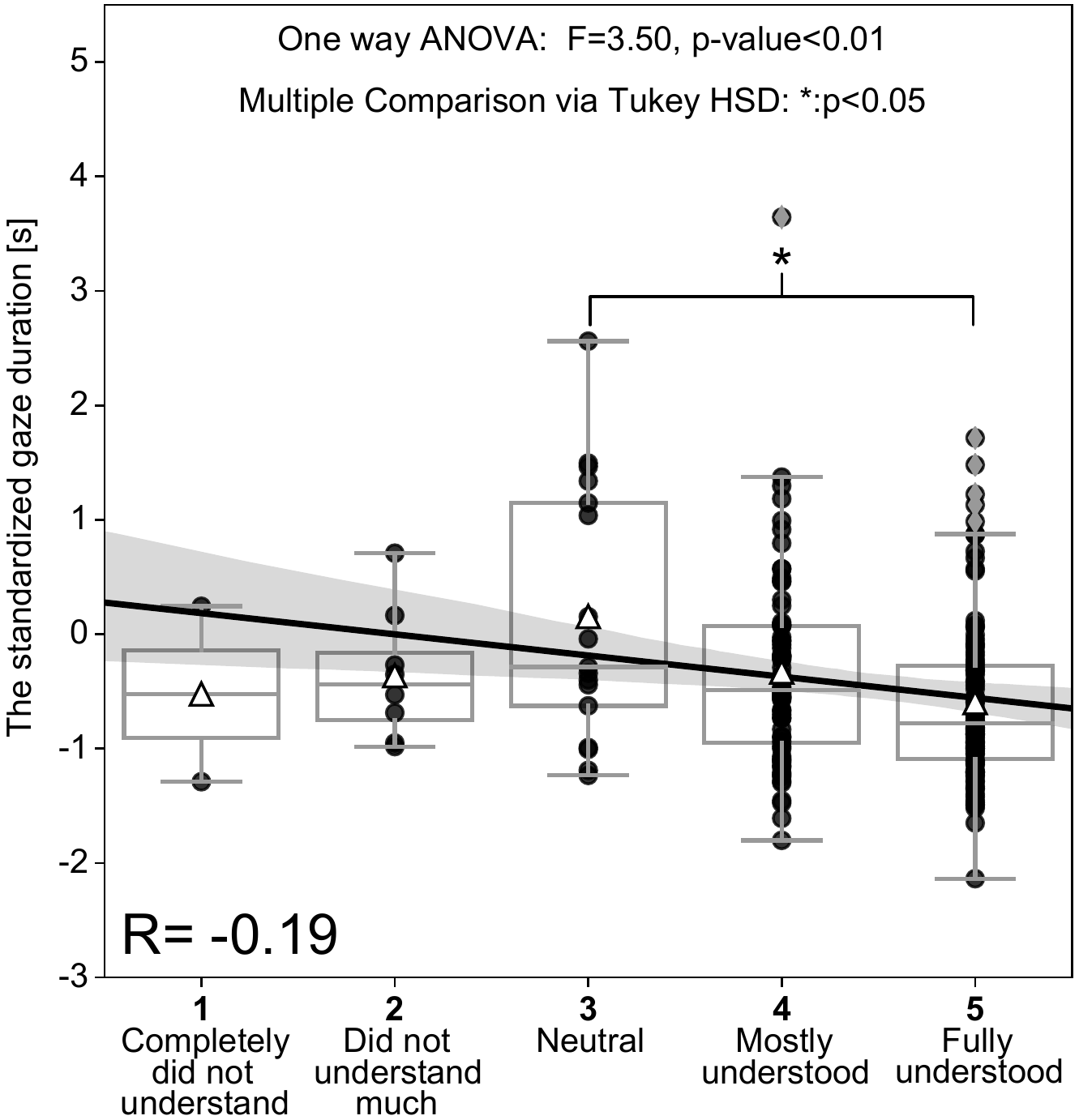}
\caption{Relationship between the understanding of the driving intentions and the gaze durations of the ten participants when interacting with the MV.}
\label{fig:md_cor}
\end{minipage}
\hspace{4mm}
\begin{minipage}{0.48\linewidth}
\centering
\includegraphics[width=1\linewidth]{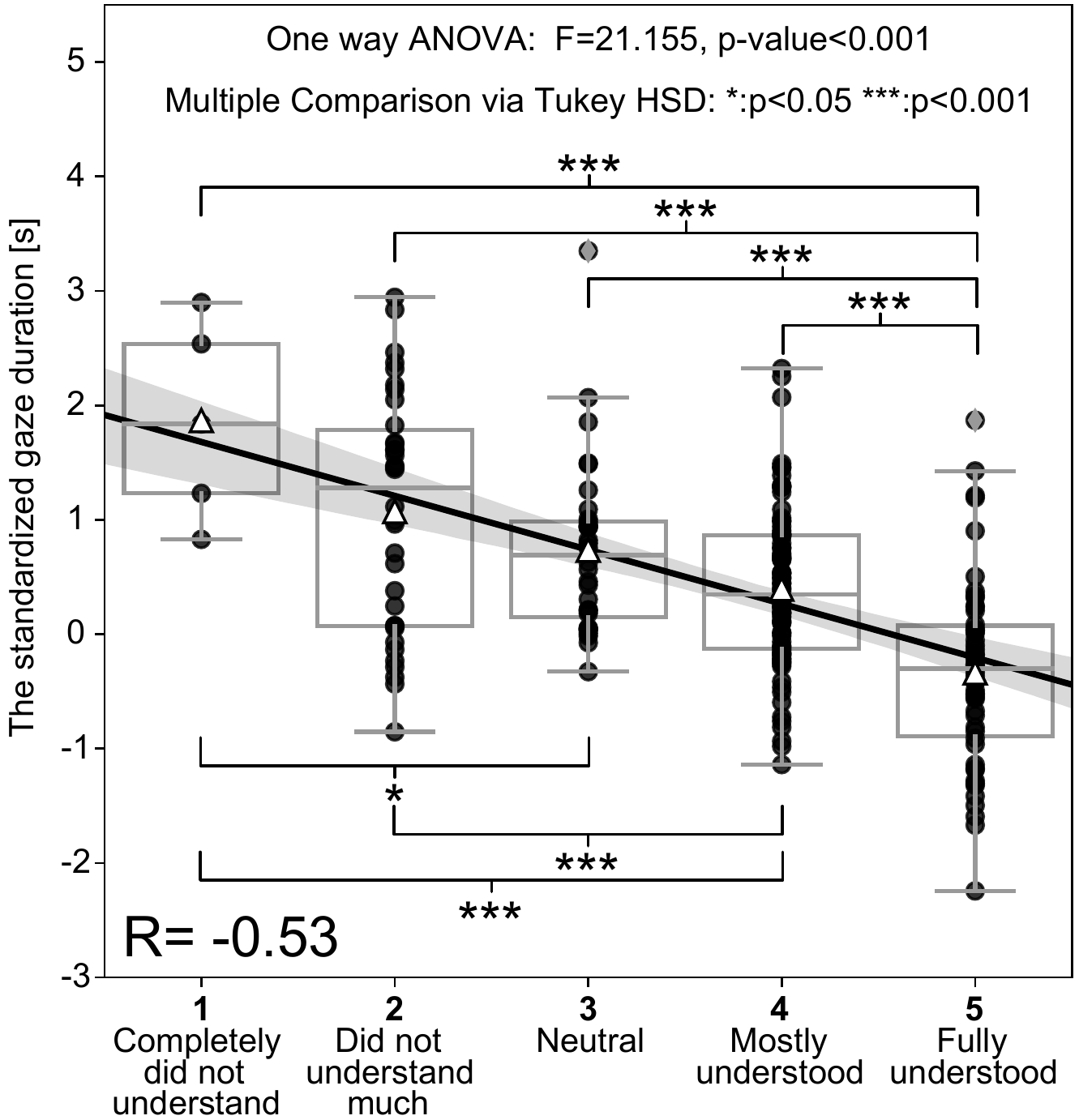}
\caption{Relationship between the understanding of the driving intentions and the gaze durations of the ten participants when interacting with the AV.}
\label{fig:ad_cor}
\end{minipage}
\vspace{-4mm}
\end{figure*}

The relationship between the participants' subjective evaluations and gaze durations on the MV and AV was analyzed by a correlation analysis.
However, there was an issue with the individual difference in how long each participant gazed at the vehicle as shown in Fig.~\ref{fig:time}.
To exclude the influence of individual difference from the correlation analysis, the mean and the standard deviation of the gaze durations of each participant were standardized to $0$ and $1$, respectively. 
Here, the gaze duration of the $n$-th participant in the $i$-th trial is written as $t^n_i\in \{T^n_{MV},T^n_{AV}\}$, where the $T^n_{MV}$ is a set of gaze durations on the MV and the $T^n_{AV}$ is a set of gaze durations on the AV.
Hence, the standardized gaze duration of the $n$-th participant in the $i$-th trial is:
\newline
\vspace{0mm}
\begin{eqnarray}
 \tilde{t}^n_i&=&\frac {t^n_i-\mu^n}{\sigma^n},\\
 \mu^n&=&\frac{\sum_{i=0}^{I^n}(t^n_i)}{I^n},\\
 \sigma^n&=&\sqrt{\frac{\sum_{i=0}^{I^n}(t^n_i-\mu^n)^2}{I^n}},
\end{eqnarray}
where, $I^n$ is the number of trials when the $n$-th participant interacted with the MV or AV, and $\mu^n$ and $\sigma^n$ are the mean and the standard deviation of $\{T^n_{MV},T^n_{AV}\}$, respectively.

After standardizing the gaze duration of each participant, the correlation between the subjective evaluations and the gaze durations was analyzed.
The results are shown in Fig.~\ref{fig:md_cor} and \ref{fig:ad_cor}.
Note that the horizontal axis represents the five-levels of evaluation and the vertical axis represents the standardized gaze duration.

Fig.~\ref{fig:md_cor} illustrates the relationship between the subjective evaluations of the understanding of the driving intentions and the gaze durations when the participants interacted with the MV.
Note that the mean values are shown by white triangles.
The mean values of the gaze durations corresponded to each evaluation were similar for the participants that interacted with the MV.
A linear regression was used to visualize the correlation between the two variables and is represented by a solid black line in Fig.~\ref{fig:md_cor}.
The gray area around the solid black line represents the 95\% confidence interval for the regression.
The larger the gray area, the less accurate the linear regression is and vice versa.
The confidence intervals around options {\it \ctext{1} Completely did not understand} and {\it \ctext{2} Did not understand much} were significantly larger than those around options {\it \ctext{4} Mostly understood} and {\it \ctext{5} Fully understood}.
This is because participants rarely chose options {\it \ctext{1} Completely did not understand} and {\it \ctext{2} Did not understand much}.
This also means that the participants thought it was difficult for them to understand the driving intentions. 
Meanwhile, the results of the correlation analysis showed that the correlation coefficient between the two variables was $-0.19$. In other words, there was a very weak negative correlation.

Here, a one-way ANOVA~\cite{st1989analysis} was used to verify the difference in the gaze duration corresponding to each subjective evaluation.
The null hypothesis for the ANOVA was that there were no significant differences among the groups in gaze durations.
The result of the ANOVA showed that the F-value was $3.50$ and the p-value was less than 0.01.
Therefore, the null hypothesis was rejected, and there were significant differences among the groups in gaze durations.
After that, multiple comparison via Tukey HSD was used to verify the significant differences between each pair of groups.
The result of the multiple comparison via Tukey HSD~\cite{abdi2010tukey} showed that only the pair of {\it \ctext{4} Mostly understood} and {\it \ctext{5} Fully understood} had significant differences ($p<0.05$).

The above results demonstrate that there was little negative correlation between the ten participants' understanding of the MV's driving intentions and the gaze durations when they interacted with it.
This result of MV is contrary to the hypothesis, even though the hypothesis is formulated only for the AV.
In the present study, it is difficult to rigorously verify the reason for this. However, some indications can be gained from the interviews conducted after the experiment.
Most of the participants reported that they did not pay much attention to the MV because the driver (experimenter) was sitting in it, and the driver would avoid them even if they did not understand the vehicle's intentions.
This indicates that the trust the participants had toward the driver affected the interaction.

Fig.~\ref{fig:ad_cor} shows the relationship between the subjective evaluations of the understanding of driving intentions and the gaze durations when the participants interacted with the AV.
This figure shows that as the ten participants' understanding of the AV's driving intentions became accurate, they spent less time observing the AV.
The correlation coefficient between the two variables is $-0.53$.
The 95\% confidence interval of linear regression is also smaller than in the case of the MV. This indicates that the accuracy of regression is higher than in the case of the MV.
This suggests that the participants' gaze duration on the AV was related to their understanding of the AV's driving intentions.

The same process was used for statistical analysis of the AV data that was used for analyzing the MV data.
A one-way ANOVA was used to verify the significant difference among the groups of gaze durations on the AV. 
Meanwhile, the multiple comparison via Tukey HSD was used to verify the significant difference between each pair of groups.
The result of the ANOVA showed that the F-value was $21.15$ and the p-value was lower than $0.001$.
It showed that there were significant differences among the groups in gaze durations on the AV.
Meanwhile, the test results of the multiple comparison showed that there was a significant difference between each pair of groups except for the pair of {\it \ctext{1} Completely did not understand} and {\it \ctext{2} Did not understand much}, the pair of {\it \ctext{2} Did not understand much} and {\it \ctext{3} Neutral}, and the pair of {\it \ctext{3} Neutral} and {\it \ctext{4} Mostly understood}.

The results above validate the hypothesis that if a pedestrian does not accurately understand the driving intentions of an AV, then their gaze duration will increase during the interacting compared to when they understand the driving intentions.

\section{CONCLUSION}
This paper presented a study regarding measurable behaviors of a pedestrian who has various levels of understanding of the driving intentions of an AV in an interaction.
We formulated and proposed a hypothesis based on situation awareness that if a pedestrian did not clearly understand the driving intentions of the AV, then their gaze duration on the AV would be longer during the interaction than if they understood the intentions clearly.
To verify this hypothesis, a robotic wheelchair was used in a pedestrian--vehicle interaction experiment.
This vehicle could be driven manually, or it could automatically drive following predesigned routes.
The participants' gaze information and subjective evaluation of their understanding of the driving intentions were collected.
The following conclusions were established:
\begin{enumerate}
\item When the participants interacted with the AV, their gaze durations on the AV were negatively correlated with their subjective evaluations of understanding the AV's driving intentions, but this was not the case with the MV.

  \item The participants perceived the driving intentions of the AV as more difficult to understand than the driving intentions of the MV when they interacted with the vehicle. 
  \item Most of the participants' gaze durations on the AV were longer than on the MV. 
\end{enumerate}
From the above results, it could not be ruled out that the participants' gaze durations included the observation time generated by their curiosity about the AV.
However, the participants' gaze durations could still be negatively correlated with their subjective evaluations of the understanding.
This illustrates the fact that participants increased their gaze duration on the AV when not understanding AV's driving intentions.

After analyzing the experiment, the answer to the question ``What gaze behavior do pedestrians display in interactions when they do not understand the intention of an automated vehicle?'' is that the pedestrians continue to observe the AV.
We consider that the gaze duration on the AV by pedestrians could be used to determine whether pedestrians understand the AV or not.
Besides, we also consider that this gaze behavior indicates that pedestrians need more information from the AV about its driving intentions. 
This could also be considered a breakthrough point for resolving the timing of providing information to pedestrians by the AV.
Thus, two recommendations are proposed to the designers of eHMI:
\begin{enumerate}
  \item When a pedestrian is engaged in an interaction with the AV, the driving intentions of the AV should be provided.
  \item If the pedestrian still gazes at the AV after the AV displays its driving intentions, the AV should provide more clear information about its driving intentions.
\end{enumerate}

In future studies, we will develop a computational model to predict whether a pedestrian understands AV's driving intentions by using pedestrian's gaze duration.
Meanwhile, an evaluation system will be established to evaluate whether the designed eHMI is easily understood.


\bibliographystyle{ieeetr}
\bibliography{sample-base}

\end{document}